\documentclass[sigconf, nonacm]{acmart}
\AtBeginDocument{%
  \providecommand\BibTeX{{%
    \normalfont B\kern-0.5em{\scshape i\kern-0.25em b}\kern-0.8em\TeX}}}

\usepackage{CJK}

\begin{document}
\begin{CJK*}{UTF8}{gbsn}

\title{GraV: Grasp Volume Data for the Design of One-Handed XR Interfaces}

\author{Alejandro Aponte}
\authornote{Both authors contributed equally to this research.}
\email{aponte@ucsb.edu}
\orcid{0000-0003-2122-0451}
\affiliation{%
  \institution{University of California}
  \city{Santa Barbara}
  \state{CA}
  \country{USA}
}

\author{Arthur Caetano}
\authornotemark[1]
\email{caetano@ucsb.edu}
\orcid{0000-0003-0207-5471}
\affiliation{%
  \institution{University of California}
  \city{Santa Barbara}
  \state{CA}
  \country{USA}
}

\author{Yunhao Luo}
\email{yunhaoluo@ucsb.edu}
\orcid{0009-0004-6219-8021}
\affiliation{%
  \institution{University of California}
  \city{Santa Barbara}
  \state{CA}
  \country{USA}
}

\author{Misha Sra}
\email{sra@ucsb.edu}
\orcid{0000-0001-8154-8518}
\affiliation{%
  \institution{University of California}
  \city{Santa Barbara}
  \state{CA}
  \country{USA}
}

\renewcommand{\shortauthors}{Aponte \& Caetano, et al.}
\newcommand{\pname}{GraV}

\newcommand\change[1]{#1}

\begin{abstract}

Everyday objects, like remote controls or electric toothbrushes, are crafted with hand-accessible interfaces. Expanding on this design principle, extended reality (XR) interfaces for physical tasks could facilitate interaction without necessitating the release of grasped tools, ensuring seamless workflow integration. While established data, such as hand anthropometric measurements, guide the design of handheld objects, XR currently lacks comparable data, regarding reachability, for single-hand interfaces while grasping objects. To address this, we identify critical design factors and a design space representing grasp-proximate interfaces and introduce a simulation tool for generating reachability and displacement cost data for designing these interfaces. Additionally, using the simulation tool, we generate a dataset based on grasp taxonomy and common household objects. Finally, we share insights from a design workshop that emphasizes the significance of reachability and motion cost data, empowering XR creators to develop bespoke interfaces tailored specifically to grasping hands.

\end{abstract}



\begin{teaserfigure}
  \includegraphics[width=\textwidth]{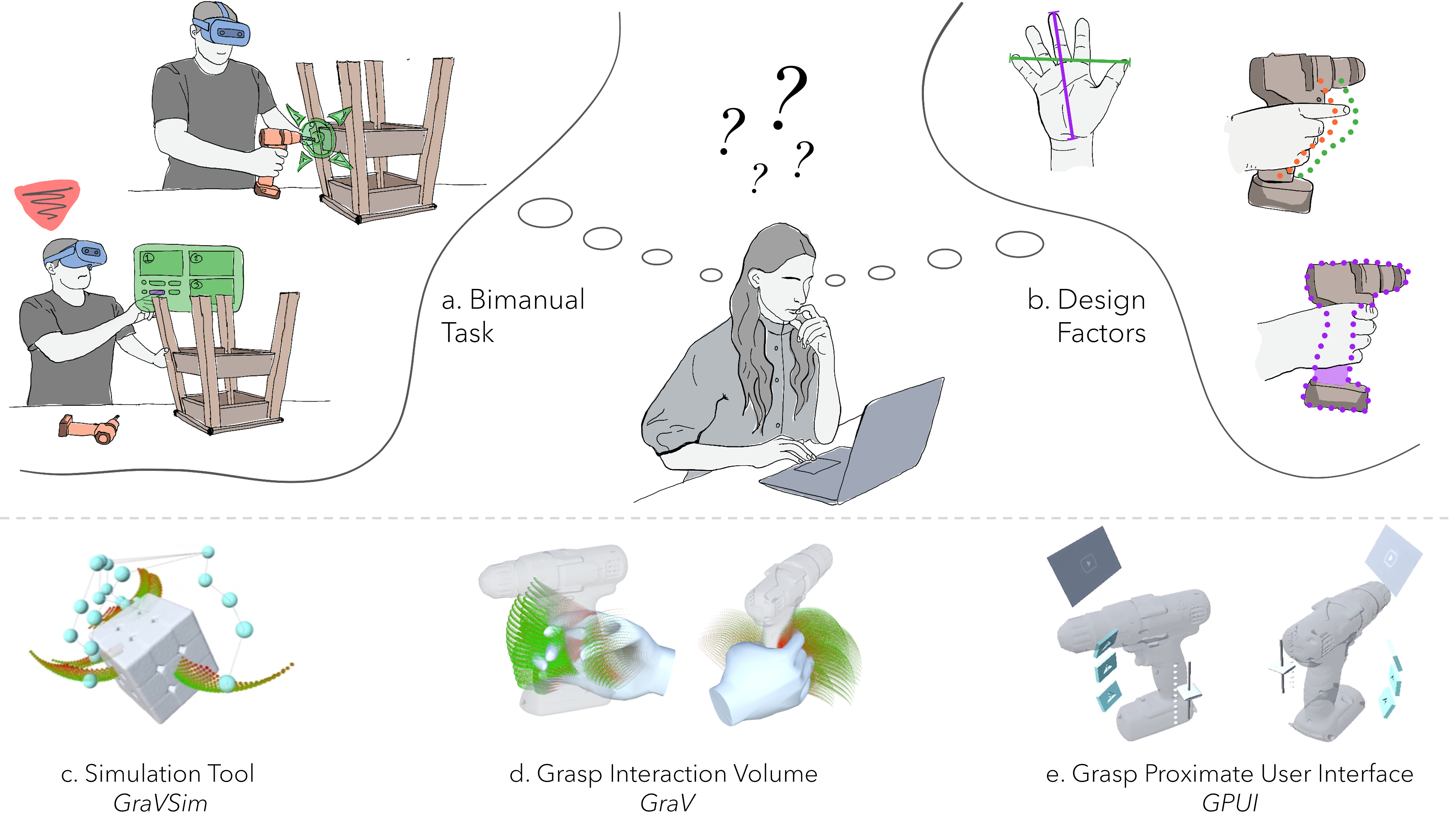}
  \caption{
  (a) When performing a bimanual task, users must place down the manipulated object to interact with mid-air interfaces, causing inconvenience and disrupting task flow. (b) Designers approaching this problem face a complex combination of parameters including hand size, motion range, and object shape. (c) GraVSim facilitates this process by simulating hand-object interactions to create GraVs (d), representing reach and motion cost. (e) Our tools simplify the design process of GPUIs, interfaces that offer finger-accessible interactions while maintaining object grasp.}
  \Description{At the center of the figure is a designer at the laptop with two thought balloons and interrogation marks on their head. The top-right balloon is labeled ``a.Bimanual Task'' and shows a person fixing a stool with a drill while wearing an MR headset showing guidance elements. The user seems annoyed when they have to put down the drill to interact with the mid-air interface. The top-left is labeled ``b.Design Factors'' balloon shows a hand with length measurements, a hand holding a drill with dashed lines at the edge of the space the finger can reach, and an outline around the shape of the drill. In the row below the previously described figure, there are three simulation figures. The first element is labeled ``c. simulation tool GraVSim'' and shows an armature holding a cube with a point cloud around it. The point cloud colored points in a gradient from green to red depicting the different motion cost values. The second element is labeled ``d. Grasp Interaction Volume'' and shows two perspectives of a hand holding a drill, both with a color-coded point cloud around. The third and last element on the bottom-right is labeled ``e. Grasp Proximate User Interface GPUI'' and shows an AR interface around the drill with three buttons, a slider, and a video player. All of the UI elements seem accessible by the fingers.}
  \label{fig:teaser}
\end{teaserfigure}

\maketitle

\section{Introduction}

Physical tasks like assembly, repair, or surgery often require the use of both hands. One hand holds a tool such as a screwdriver or a scalpel, while the other hand manipulates the object being worked on. To assist in these types of tasks, many devices feature hand-accessible interfaces while the user is grasping the object. These hand-accessible interfaces facilitate interaction with the system while enabling a more continuous workflow without releasing the device. XR interfaces have the potential to apply this strategy beyond the current form factor and intended use of the grasped object, by overlaying virtual UI elements surrounding the hand, irrespective of the grasped object's original design. 

There has been a growing interest in the design of extended reality (XR) applications for physical task assistance \cite{palmarini2018systematic, henderson2007augmented, westerfield2015intelligent}. Nonetheless, current XR interfaces are not well suited for tasks that occupy the user's hands. Researchers have proposed hands-free interaction modalities such as gaze and speech which have been explored successfully in XR (see \cite{monteiro2021hands,plopski2022eye} for recent surveys). However, these approaches still have some limitations. Prior work shows that gaze-based interfaces are not ergonomic and introduce newer challenges such as neck fatigue, slower interactions with more errors, and reduced spatial awareness, while speech-based interfaces are challenging to use in loud environments, have high latency, and have difficulty interpreting different accents \cite{monteiro2021hands,plopski2022eye}. For head-mounted displays (HMDs), mid-air interactions are a common input modality. However, these interactions can be uncomfortable and cause fatigue, often referred to as the ``gorilla arm effect,''~\cite{evangelista2021xrgonomics,faleel2021hpui,dudley2019performance,harrison2012body}. Additionally, regardless of XR interaction techniques, the corresponding virtual interfaces often tend to be flat-screen projections with point-and-click controls, largely inspired by desktop or mobile 2D user interfaces. These 2D-inspired approaches disregard critical ergonomic factors, such as hand or finger reach and user pose \cite{laviola20173d}. 

Hands are our primary means of physical interaction, both with the world and with XR. Placing UI elements on or near the hand can leverage our familiarity and proprioception to enable the design of comfortable and easy-to-use interfaces \cite{harrison2011omnitouch, tung2015user}. When designing XR interfaces, prior work has shown on-hand designs to be more subtle, less socially awkward \cite{laviola20173d}, and less tiring \cite{dudley2019performance,koutsabasis2019empirical} than other forms of XR interactions. However, the scenario of a user grasping a physical object in one hand, with the interface reachable by the same hand is still unexplored. In our work, we focus on XR one-handed interfaces reachable by one hand while grasping an object, without extending the arm, and interactable with partial or complete hand or finger movements depending on the grasp. We refer to them as Grasp-Proximate User Interfaces (GPUIs).

The design of these interfaces, meant to be used by an occupied hand, requires a basic understanding of the hand-object interaction and how users manipulate their hands to engage with the interface. Creating one-handed interfaces for users engaged in physical tasks that involve tools or objects presents several challenges. Designers must consider how users grip objects (e.g., power grip for using a drill), which fingers have partial or full movement for interacting with the UI elements (e.g., thumb may have partial movement while holding a drill), and what are the reachable areas based on grasp. To address this, and assist designers in the creation of GPUIs, we defined a series of design factors and a design space that represents the hand-object interaction based on function, reachability, motion cost, and emerging boundaries. We then implement these design factors through a Unity simulation tool (GraVSim) to generate data that provides information about the hand segments' motion, reachability, and cost in the form of a three-dimensional space called \underline{Gra}sp Interaction \underline{V}olume(\pname). To facilitate access to \pname~data we share a dataset of GraVs that we generated using a combination of grasps (as defined by the taxonomy) and everyday objects in the YCB Affordance dataset~\cite{corona2020ycb_affordance}. 

We believe that \pname~can assist designers in the development of grasp-proximate user interfaces by providing a better understanding of the hand-object coupled interaction through quantifiable data and tangible boundaries. GraV does not prescribe a GPUI design; instead, it aims to inform design decisions by providing reach and interaction cost data. To explore this, we conducted a workshop and formative group interviews with XR designers from various backgrounds, where they created a series of grasp-proximate user interfaces for objects in a cooking scenario. The designers' explorations and key insights for the use of GraV in their workflows as well as future improvements of the GraVSim tool are presented as part of this work. 

The proposed work builds on prior research on thumb-to-finger interactions \cite{trudeau2012thumb,wong2018fingert9}, microgestures \cite{soliman2018fingerinput, sharma2019grasping,sharma2021solofinger, joshi2023transferable}, on the hand interfaces \cite{faleel2021hpui} and explorations using physical objects as part of the interaction design \cite{he2023ubi, jain23ubitouch}. Yet, even with the increasing interest in making XR interfaces more practical for everyday real-world interactions, addressing the physical elements of these interactions is still a challenge for developers \cite{ashtari2020creating}. This is due to the lack of transferable design guidelines and technical solutions that provide quantifiable data to assist designers, particularly for grasp-proximate interfaces. The importance and value of these systems are emphasized by \cite{evangelista2021xrgonomics}, where they explore this approach by providing ergonomic data to optimally place mid-air interfaces within the proximity of the user body. We believe that the approach will help provide designers with resources to assist in breaching the gap between the design process and XR interfaces that can adapt physical tasks by facilitating the understanding of hand-object coupled interaction for grasp-proximate user interfaces. We present the following contributions:

\begin{itemize}
    \item A set of design factors that represent the hand-object coupled interaction, and a design space comprised of the three fundamental variables always present in the design of one-handed grasp-proximate user interfaces (GPUIs) in XR.
    \item A downloadable Unity package (GraVSim) for designers and developers to create customized GraVs using our parameterized 3D hand model.~\footnote{\url{https://github.com/HAL-UCSB/grav-sim}}
    \item A GraV dataset, containing reachability and cost data as a point cloud, based on standard anthropometric values, grasp taxonomy, and common household objects. 
    \item An evaluation of how GraV can be used to develop GPUIs through a workshop and formative interviews with XR designers. The workshop demonstrated how designers can approach the design of these types of interfaces using simulated quantifiable data, how they would incorporate this data in their workflows, and future recommendations to improve the simulation tool (GraVSim)

\end{itemize}

\section{Related Work}\label{sec:rw}

\subsection{Design of Single-Handed Interfaces}
To clarify how GPUIs and \pname~relate to prior work, we present an overview of previous research on the design of interfaces with support to one-handed interactions accessed by hand motion. This space has been previously explored by Faleel et al. \cite{faleel2020user}, who proposed a framework for hand-proximate user interfaces (HPUIs) which are virtual interfaces registered to the user's hand or the space around it. Sharma et al.~\cite{sharma2019grasping} elicited single-hand microgestures for handheld objects and proposed three categories of gestures according to their action location: In-Air, On-Body, and On-Object. Ergonomics of single-handed interactions have been studied, pointing out that factors such as interface orientation, size of handheld object, and arm pose can affect physical comfort and performance~\cite{xu2018hand, trudeau2012thumb, evangelista2021xrgonomics}.

\subsubsection{In-Air}
In-air single-handed interactions use the available space around the hand to support interaction modalities such as XR interfaces~\cite{faleel2020user, perella2023evaluating, kim2022pseudo, xu2018hand} and gestures~\cite{faleel2020user, sharma2021solofinger}. HPUIs that support in-air interactions can follow a layout that uses the space above the hand to show elements such as a panel of icons~\cite{faleel2020user}. Kim et al.~\cite{kim2022pseudo} compared in-air typing in VR with pseudo- and self-haptics alternatives, finding they support comparable typing performance but underperform in user experience and preference.

These interfaces can support in-air interactions with head-mounted displays in a single-hand-free setting, but their usability is compromised when the user grasps an object that obstructs finger motion. \pname~mitigates this problem by providing information about the space reachable by the fingers while grasping an object. Designers of these types of interfaces must decide on the layout of elements, which is not a trivial decision given there is limited information about the comfort of hand interactions during the design phase.~\pname~tackles this by providing joint rotation costs in a point cloud reachable by users' fingers. Moreover, when the user's hand diverges too much from the assumed design parameters, due to anatomic differences or motor disease, these interfaces can become unusable. ~\pname~can be personalized for each user to reflect their individual interaction space, allowing designers to tailor their interfaces to the interaction space of a specific user.

In SoloFinger, researchers demonstrated a method to design microgestures that are robust to false activation based on the analysis of movement signatures, enabling users to perform gestures reliably while holding objects~\cite{sharma2021solofinger}. However, choosing among the possible microgesture options while considering different grasps and different objects can be a challenging task. \pname~can be used by designers to identify the most suitable microgesture based on available movement space and pose costs.

Xu et al. \cite{xu2018hand} evaluated comfort metrics of interactions in the range of motion of the wrist and found physical comfort, pointing speed, and pointing accuracy to be higher for interfaces with fixed orientations. In more recent work, Evangelista et al.~\cite{evangelista2021xrgonomics} proposed a toolkit to visualize the interaction cost of mid-air interactions, supporting the design of ergonomic XR interfaces. XRgonomics divides the mid-air interaction space into voxels that are accessible by a simplified arm structure using inverse kinematics. Different ergonomic costs can be assigned to the voxels. These projects can provide designers with valuable data about in-air interactions; however, they do not offer information about finger motion and do not consider motion obstructions caused by a grasped object or self-collision. To support the design of interfaces for hands grasping an object, we propose \pname, a volume that represents the space reached by the fingers of a grasping hand.

\subsubsection{On-Hand and On-Body}

On-hand single-handed interactions use the surface of the hand to enable interaction modalities such as XR interfaces~\cite{faleel2020user, perella2023evaluating, kim2022pseudo, whitmire2017digitouch}, on-body projected interfaces~\cite{yamamoto2007palmbit, harrison2012body, mistry2009wuw, harrison2011omnitouch}, gestures~\cite{huang2016digitspace, soliman2018fingerinput}. One key benefit of on-hand interfaces is the tactile feedback when the hand touches itself, commonly in finger-thumb contact. Another benefit of on-hand interfaces is the potential to leverage proprioception for eyes-free interaction~\cite{huang2016digitspace}.

HPUIs can support on-hand interactions by showing virtual interface elements on the hand and support several gestures and finger-thumb interactions~\cite{faleel2020user, perella2023evaluating}. Virtual keyboard interfaces have been deployed as on-hand interactions where keys are pressed with finger-thumb gestures~\cite{kim2022pseudo}. DigiTouch~\cite{whitmire2017digitouch} is a glove-based input device that supports thumb-to-finger interactions targeted to head-mounted displays.

Another strategy to support on-hand interactions in single-handed interfaces is on-body projected interfaces. PALMbit \cite{yamamoto2007palmbit} is an ONPI that uses a shoulder-worn projector and camera that uses the palm of the hand as an interface. SixthSense \cite{mistry2009wuw} is another example of ONPI powered by computer vision that allows the hands to be tracked for input. OmniTouch \cite{harrison2011omnitouch} has a similar shoulder-mounted projector that enables true touch inputs through depth sensing. DigitSpace\cite{huang2016digitspace} supports interaction with buttons, touchpads, and sliders based on thumb-to-finger motions. Soliman et al.~\cite{soliman2018fingerinput} proposed a design space of thumb-to-finger microgestures accompanied by a recognition system powered by depth sensing.

These interfaces are effective for projected on-hand interactions in a setting where one hand is free. However, their performance becomes uncertain when the user holds an object that restricts finger movement. To address this issue, \pname~offers insights into the reachable space for fingers. Designers of such interfaces face the challenge of determining the layout of elements, a non-trivial decision due to limited information on comfort during the design phase. \pname~addresses this by providing joint rotation costs within a point cloud that can be accessed by users' fingers. Furthermore, when users' hand positions deviate significantly from the assumed design parameters, whether due to anatomical differences or motor disorders, these interfaces may become less effective. \pname~offers the potential for personalization, allowing each user's individual interaction space to be considered.

\subsubsection{On-Object}
On-object single-handed interactions use the surface of the hand to enable interaction modalities such as XR interfaces~\cite{hettiarachchi2016annexing, he2023ubi, zhou2020gripmarks}, wearable controllers~\cite{lyons2004twiddler, thomas2002does, sturman1994survey, boring2009scroll}, on-body projected interfaces~\cite{yamamoto2007palmbit, harrison2012body, mistry2009wuw, harrison2011omnitouch}, and microgestures~\cite{sharma2021solofinger}. An additional benefit of on-object interfaces is the tactile feedback provided by the contact with the object.

The geometric alignment between a physical object and the virtual interface has been explored as one of the key factors that enable on-object in single-handed interfaces. In Annexing Reality~\cite{hettiarachchi2016annexing}, researchers matched virtual objects to physical proxies based on their shape similarity supporting haptic sensation for virtual objects. In UbiEdge~\cite{he2023ubi} authors explore the use of edges, ubiquitous geometric features, to opportunistically place virtual interfaces. 

While these interactions allow on-object interactions with opportunistic haptics, they overlook the ability of the user's hand to access different regions of the object when grasping with only one free hand. The grab interaction volumes we propose contain the surface of the object that is accessible by the user's hand while grasping an object. \pname~focuses on the reachable space around a grasp, to help expand interface design possibilities while enabling the potential for opportunistic haptic feedback \cite{fang2021retargeted,kim2022pseudo} both from thumb-to-finger interactions and from finger-to-grasped object interactions.

Gripmarks~\cite{zhou2020gripmarks} proposed the use of templates of grips, according to a proposed classification, and basic shapes of handheld objects to identify runtime opportunities for haptics. In a proof-of-concept demonstration, authors showcased the use of Gripmarks to create UIs aligned to an object surface that support tap, swipe, and gesture interactions. Even though Gripmarks supports grip-based opportunistic haptics in runtime, allowing users to interact with the system without letting go of the object in hand, it offers limited support during the UI design phase, primarily focusing on hand grip templates and object primitive shapes. \pname~complements this approach by providing designers with finger motion data as an interaction volume point cloud that takes into consideration the limitations of finger range of motion (RoM) and arbitrary grasped object geometry.

Iconic wearable on-body interfaces include devices such as a one-handed keyboard attached to the palm~\cite{lyons2004twiddler}, a wrist-bound touchpad \cite{thomas2002does}, and glove and fingertip-based input systems \cite{sturman1994survey}. \pname~can support the design of on-body interfaces by making designers aware of the space that can be reached by the fingers in a free-hand setting or while holding a physical device.

Trudeau et al.~\cite{trudeau2012thumb} studied motor performance during single-handed mobile phone use and found that smaller phones lead to better thumb performance in adduction-abduction movements.

\subsection{Hand-Object Interaction Datasets}

Hand-object interaction is a subject of interest in various fields including motion reconstruction, human-robot interaction, and action recognition~\cite{fan2023arctic, chao2021dexycb,hasson2019learning}. To assist research in this area, multiple datasets have been constructed. ARTIC is a dataset of bimanual manipulation of articulated objects like laptops and scissors and contains contact information, 3D hand and object meshes~\cite{fan2023arctic}. The DexYCB dataset contains multiview RGB-D frames of 10 subjects grasping 20 objects and provides ground-truth hand and object 3D poses~\cite{chao2021dexycb}. The ObMan dataset is a large-scale synthetic dataset of hand-object manipulation scenarios~\cite{hasson2019learning}. It comprises a total of $2772$ 3D models across eight daily life object categories. The grasps in this dataset were generated using GraspIt, a robotic grasp simulation software~\cite{miller2004graspit}. The GRAB dataset includes complete body motions recorded using motion capture markers from ten participants engaging with and grasping 51 different objects. It contains 3D hand and object positions, along with binary object contact maps~\cite{taheri2020grab}. While these datasets provide rich data for research on hand-object grasps, none of them contain grasp interaction volumes. The~\pname~dataset fills this gap and provides 367 grasp interaction volumes to support the design of GPUIs in XR. Our dataset builds on top of the YCB Affordance~\cite{corona2020ycb_affordance} dataset and contains 58 objects from the YCB object set~\cite{berk2015ycb} manually annotated with hand poses for each of the 33 types of grasps in the GRASP taxonomy~\cite{feix2015grasp}. Since not all the objects can be grasped with every grasp type, there are 367 valid combinations in total. Since the YCB dataset provides object poses and representative grasps, we use it as input data for the generation of our volume dataset, which contains 367 grasp interaction volumes generated from the valid object-grasp combinations in the YCB Affordance dataset.

Prior research has developed tools to support hand-object interaction data collection. One such tool is ARnnotate~\cite{qian2022arnnotate}, which assists users in performing various hand poses while the system records 3D hand positions, 3D object bounding boxes, images, and additional metadata. Tools like ARnnotate could be utilized to gather data for expanding the \pname~dataset we provide or for generating personalized \pname~data in real-time.

\section{Influencing Factors and Design Space}\label{sec:factors}

In our daily lives, we often interact with physical objects with interfaces optimized for single-handed use. This single-handed interaction is a familiar and comfortable strategy that is already used in the design of
common objects, such as TV remote controls, power drills, and computer mice. They capitalize on common interactions and poses that are familiar to the user to reduce the learning curve and provide comfort for prolonged use. Reachability can be optimized for supporting new functionality or expert users. For example, a gaming mouse has more buttons, but they are all reachable by the same fingers and similar grasp used for a non-gaming two-button mouse. Acknowledging this type of single-handed interaction (where the user grasps an object and interacts with an interface using the same hand) as a practical, comfortable, and familiar strategy for interacting with objects, we draw inspiration from this modality to formulate the factors that influence the design of interfaces reachable by a hand while grasping an object. In this work, we refer to this particular interface design scenario as GPUIs. We use the following definition of grasp and interchangeably use the word grip to mean the same thing. ``A grasp is every static hand posture with which an object can be held securely with one hand, irrespective of the hand orientation.''~\cite{feix2015grasp}. 

In our exploration, we segmented the hand's functions and its interaction with the grasped object to outline what parts of the hand could be used for interaction vs grasping (Section~\ref{sec:hand_segmentation}). Based on that we proceeded to identify the information needed to characterize the finger motion and its capacity for interaction (Section~\ref{sec:riv}). Through this characterization, we identified two boundary conditions that occur due to the object's intersection of the hand's space and the finger's reach. We also illustrate how the hand object itself could be used for opportunistic haptics. Finally, we present a three-dimensional design space based on the three fundamental variables that always need to be considered for the design of GPUIs. \change{We provide these new factors based on our own observations and prior literature as a new framework to address the design of GPUIs.}

\begin{figure}[!t]
    \centering
    \includegraphics[width=1\linewidth]{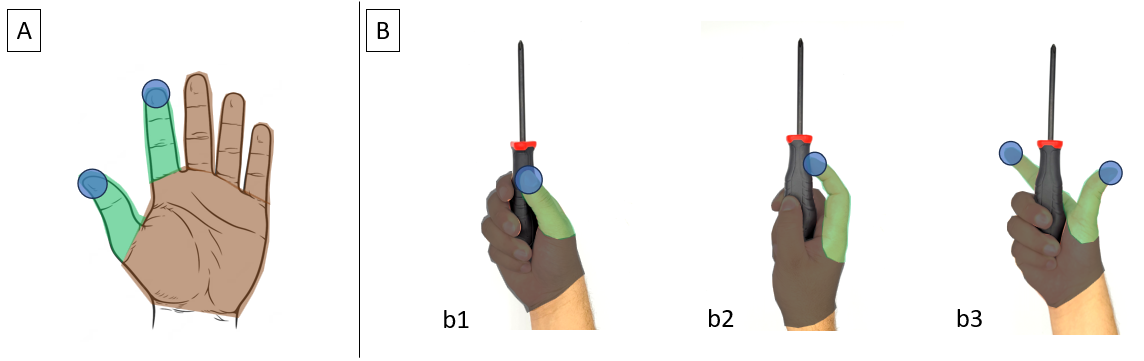}
    \caption{Hand Grasp and finger Interaction Parts. In (A), the brown region represents the part of the hand that is used for grip (i.e., the palm), the green represents the fingers that can move freely (index and thumb), and the blue dots represent the fingertips (both pads and nails) as the finger interaction parts. (b1) and (b2) represent the same grip but are adjusted to prioritize thumb and index finger mobility, respectively. (b3) shows both fingers are available to move while holding the screwdriver.}
    \Description{This picture has 2 horizontal sections. Section A shows the convention in which the paper's authors segmented the hand to reflect the parts of the hand that hold an object highlighted brown in the picture, namely the palm. It shows the fingers available to interact with the interface by highlighting them in green, namely the index finger and the thumb. Lastly, it shows the section of the finger that can interact with the surface by using blue circles over the fingertips of the index and the thumb. Section B of the picture shows 3 pictures of a hand holding a screwdriver. The first picture shows all three hand segments mentioned in picture A for the hand holding a screwdriver, where the grip pose allows the thumb to move. The tip of the thumb has a blue circle showing that the thumb fingertip was selected for the interaction part of the finger. The second picture of part B shows a similar picture, but in this case, the index can move and interact. Finally, the third picture shows the exact condition of a hand holding a screwdriver, but in this instance, both the thumb and the index finger can move. }
    \label{fig:elements}
\end{figure}

\subsection{Hand Segmentation: Grasp, Motion, and Interaction Parts}
\label{sec:hand_segmentation}
In this section, we present three categories we formulated to refer to distinct parts of the hand when grasping an object, with a specific focus on finger-based interactions. A fundamental aspect of the interaction between the hand and the object pertains to the equilibrium between ensuring grasp stability to prevent the device from falling or losing appropriate contact and allocating fingers for interacting with the intended interface. Everyday scenarios, such as smartphone usage prioritizing one-handed thumb access or using a video-game controller to balance the grasp and finger reach to incorporate speed and precision, exemplify this equilibrium. In addition, while the hand retains its hold on the object, depending on the intended interaction, the grasp can be relaxed to allow for more reach by one or more fingers. This mimics the use of one-handed tools, such as adjusting the velocity or direction of a power drill. How a person grasps a physical object depends on both the object's characteristics (e.g., weight and shape) and the nature of the task being performed. Interactions with an interface add a new dimension to hand-object interactions where the UI elements and corresponding finger-based interaction techniques need to be considered. The following components divide the hand into three functional categories:

\paragraph{\textbf{Grasping Elements}} The parts of the hand that are used for grasping an object resulting in their restricted motion. They change depending on the grasp type and the object being grasped. The brown shaded part of the hand presented in Figure~\ref{fig:elements} part A provides an example of this for a hand using 3 fingers for grasping. 

\paragraph{\textbf{Unconstrained Elements}} The parts of the grasping hand that can move without compromising the grip equilibrium and the capability to perform the intended physical task with the grasped object. In Figure~\ref{fig:elements} part A, this is represented by the fingers shaded green. 

\paragraph{\textbf{Interaction Elements}} The parts of the hand and finger segments that can be used to interact with the virtual interface. Depending on the intended interaction modality and RoM, different segments of different fingers can be used to interact with the UI elements (e.g., knuckle, fingerpad, nail). Based on comfort findings in prior work \cite{huang2016digitspace}, we chose the fingertips, including both the bottom and top of the fingertips (i.e., fingerpad and nail), as a point input in our work. A representation of this is provided in Figure~\ref{fig:elements} part A shown as blue circles on the fingertips. 

When designing a GPUI for finger interaction, a designer would first need to determine the grasping elements followed by identifying the elements that can move and their RoM. Once the available fingers are identified, designers need to select their preferred finger segment(s) to interact with the interface elements or track for gesture interactions.
Figure~\ref{fig:elements} part B represents how a user can prioritize the use of either the thumb (b1) or the index finger (b2) while holding a screwdriver. Part (b3) in the same figure, represents the user choosing a grip that allows them to use both the index and thumb and their fingertips as input interaction parts. 

\begin{figure}[!t]
    \centering
    \includegraphics[width=1\linewidth]{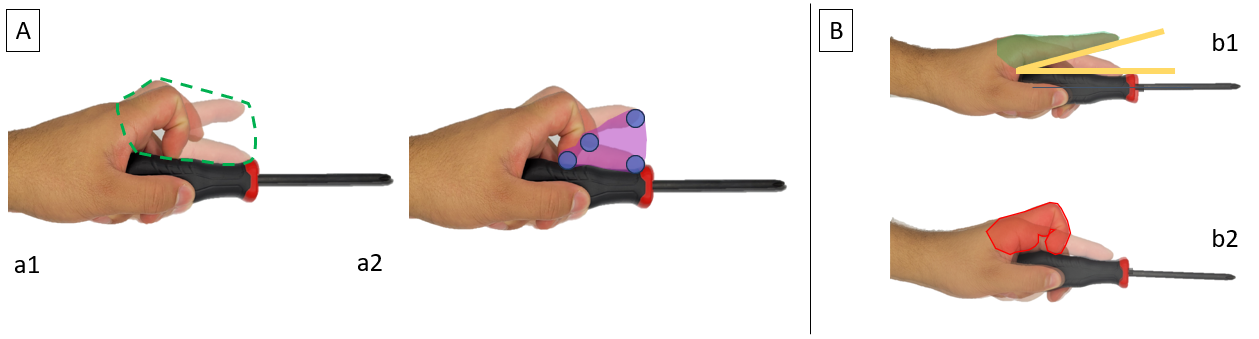}
    \caption{Hand Motion Tracking and Cost Evaluation. (a1) shows a 2D representation of the finger motion range for the index finger, while (a2) shows a 2D representation of the accessible interaction volume for the index fingertip. (b1) and (b2) represent a lower cost and a higher cost motion for the index finger, respectively.}
    \Description{Hand Motion Tracking and Cost Evaluation. On the far left (a1), we have a representation of a hand holding a screwdriver with a visual of the Finger Motion Range, and (a2) shows a 2D representation of the reachable interaction volume for the index fingertip. On the left side of the figure, there are two pictures of a hand holding a screwdriver, each representing different index finger positions with different motion costs. The high cost is painted in red, and the low cost in green.}
    \label{fig:motion}
\end{figure}

\subsection{Motion, Reach, and Displacement Cost} \label{sec:riv}
Here we focus on the finger's capacity to move and interact with the interface. To assess the interaction possibilities and limitations, we segment finger actions into three factors. The Finger Motion Range represents the motion capacity with respect to the grasped object, the Accessible Interaction Volume represents the available interaction space, and the Displacement Cost shows the effort needed to traverse the Accessible Interaction Volume, and the finger Displacement Cost. 
Each of these factors represents a concrete set of ranges that can be used to determine potential interactions. 

\paragraph{\textbf{Finger Motion Range}} The total motion space that is available to unconstrained fingers (i.e., the maximum finger RoM) when an object is grasped. The RoM gets obstructed by the object and the grasp type due to self-collision. Figure~\ref{fig:motion} part A:a1, presents an example of a finger that cannot go lower due to the surface of the screwdriver's handle. This can also be evaluated for other parts of the hand.

\paragraph{\textbf{Reachable Interaction Volume}} The motion space that can be accessed by the \textit{Interaction Elements} when holding an object. Similar to the motion range, the interaction volume changes depending on the object and grip type. It also changes depending on the selected finger segments for interaction (e.g., fingertip vs. the second digit) as seen in Figure~\ref{fig:motion} part A:a2. In that example, the blue dots represent the tracking of the fingertip as the interaction part, and the purple shade represents the \textit{Reachable} Interaction Volume.

\paragraph{\textbf{Displacement Cost}} The cost of the fingertip to reach any point in the interaction space from the resting position of the grip. Different metrics can be used to estimate the cost using factors like joint angular displacement, muscle strain, and subjective discomfort levels. In our work, we focus on joint rotation costs. Figure~\ref{fig:motion} part B shows how reaching a higher point in space with a straight index finger (b1) might be less costly than traversing the tool surface by curling the finger (b2). 

To effectively design a GPUI, it is imperative to consider the motion limitations of the hand and the reachable interaction volume when grasping an object, because the primary design element of GPUIs is to be reachable while grasping an object. The Displacement Cost can help provide a reference map of areas in the interaction space that can be used or should be avoided by developers.

\subsection{Boundaries} 
\begin{figure}[!t]
    \centering
    \includegraphics[width=1\linewidth]{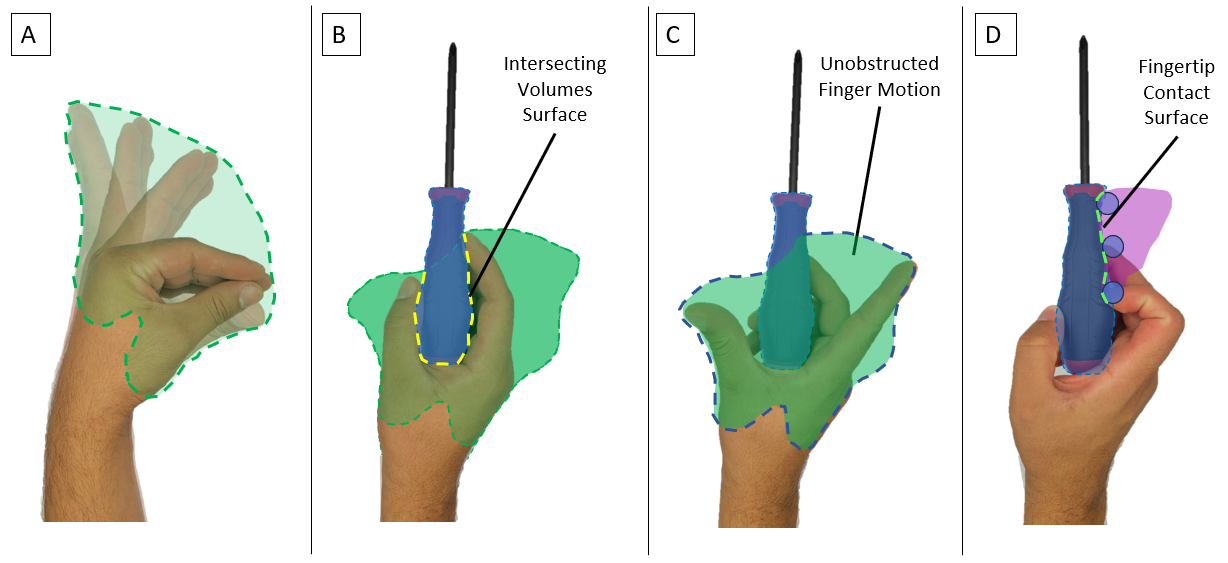}
    \caption{Object and motion boundaries. (A) shows a free-hand range of motion, and (B) shows the object boundary that is formed when an object's volume intersects with the hand's free-motion volume. (C) represents the motion boundary for both the index finger and thumb while holding a screwdriver with no external obstruction. (D) Shows the opportunistic surface haptic available for the index finger. }
    \Description{Picture with 4 sections A through D showing examples of object and motion boundaries. Picture A shows the range of motion for a free hand, creating an unobstructed volume. Picture B shows how a screwdriver's volume obstructs the free hand's volume. The intersecting surface of the 2 volumes is the object boundary. Picture C represents the motion boundary for the index finger and thumb while holding a screwdriver with no external obstruction. (D) Shows the opportunistic surface haptic available for the index finger by marking the intersection between the interaction volume and the object volume with a yellow line and blue circles representing the interaction element of the hand traversing the surface.}
    \label{fig:boundary}
\end{figure}

In this section, we define the hand as a system possessing a maximum RoM in its unobstructed or free-hand state (Figure~\ref{fig:boundary} part A), and a constrained RoM when grasping an object or in the grip state. In the free-hand state, both the palm and fingers can execute unrestricted movements (e.g., flexion, extension), with self-collision being the main limitation. However, when the hand grasps different objects, this freedom is impeded, diminishing hand motion and finger reachability. We outline two boundaries that emerge from the free- and grip-states, namely, an object boundary, which is caused by the intersecting volume of the object being grasped, and the motion boundary, which represents the maximum outward reach of the fingers. While our main focus in this work is on the fingers, other parts of the hand can be included in the evaluation of the boundaries.

\subsubsection{Object and Motion Boundaries} 
When an object is grasped, it intrudes upon the mobility space of the hand, constraining the range of motion of the fingers. It also restricts the hand's capacity to adjust the grip based on factors such as weight, size, and intended use, among others. 

\paragraph{\textbf{Object Boundary}} The part of the object that intersects the RoM of the free hand creates a boundary surface between the object and the RoM. This results in diminishing the available free-hand RoM as seen in Figure~\ref{fig:boundary} part B. 

\paragraph{\textbf{Motion Boundary}} This represents the surface of the outer trajectory of the finger's RoM that is not obstructed by the object (or self-collision of the fingers). Essentially, for each of the fingers, as they traverse the volume they can move in, the external boundary is the surface that represents their maximum unobstructed reach as presented in Figure~\ref{fig:boundary} parts C.

These two boundaries represent a dual envelope condition that surrounds the hand when grasping an object. The object boundary represents the motion limitations with respect to the object and the motion boundary represents the furthest unobstructed fingertip reach point across its motion space. The motion boundary, even though it is primarily dependent on finger reach, can be reduced by intersecting environment elements (e.g., using a screwdriver inside a tight corner in a cabinet). Knowing the environmental elements, that can intersect the \textit{Motion Boundary,} can be particularly useful when estimating the available interaction space. It is worth noting that in cases where the grasped object goes around the hand or the fingers (i.e., holding a pitcher), the pitcher would be considered as an object boundary and not an environment element. Section~\ref{sec:grav} expands on this with the Unity Tool workflow to block regions of the object that should not be used for the placement of interface elements.

\subsubsection{Opportunistic Surface Haptics} 

When we look at the intersection of the object boundary and the interaction space, we get a sub-section of the object surface that can be reached by the input region of the finger (i.e., the fingertip) Figure~\ref{fig:boundary}. These surfaces offer a unique opportunity to provide passive haptic feedback as allowed by the grip and the task, in addition to thumb-to-finger haptics \cite{fang2021retargeted, faleel2021hpui}. 

\begin{figure*}[!t]
    \centering
    \includegraphics[width=.9\linewidth]{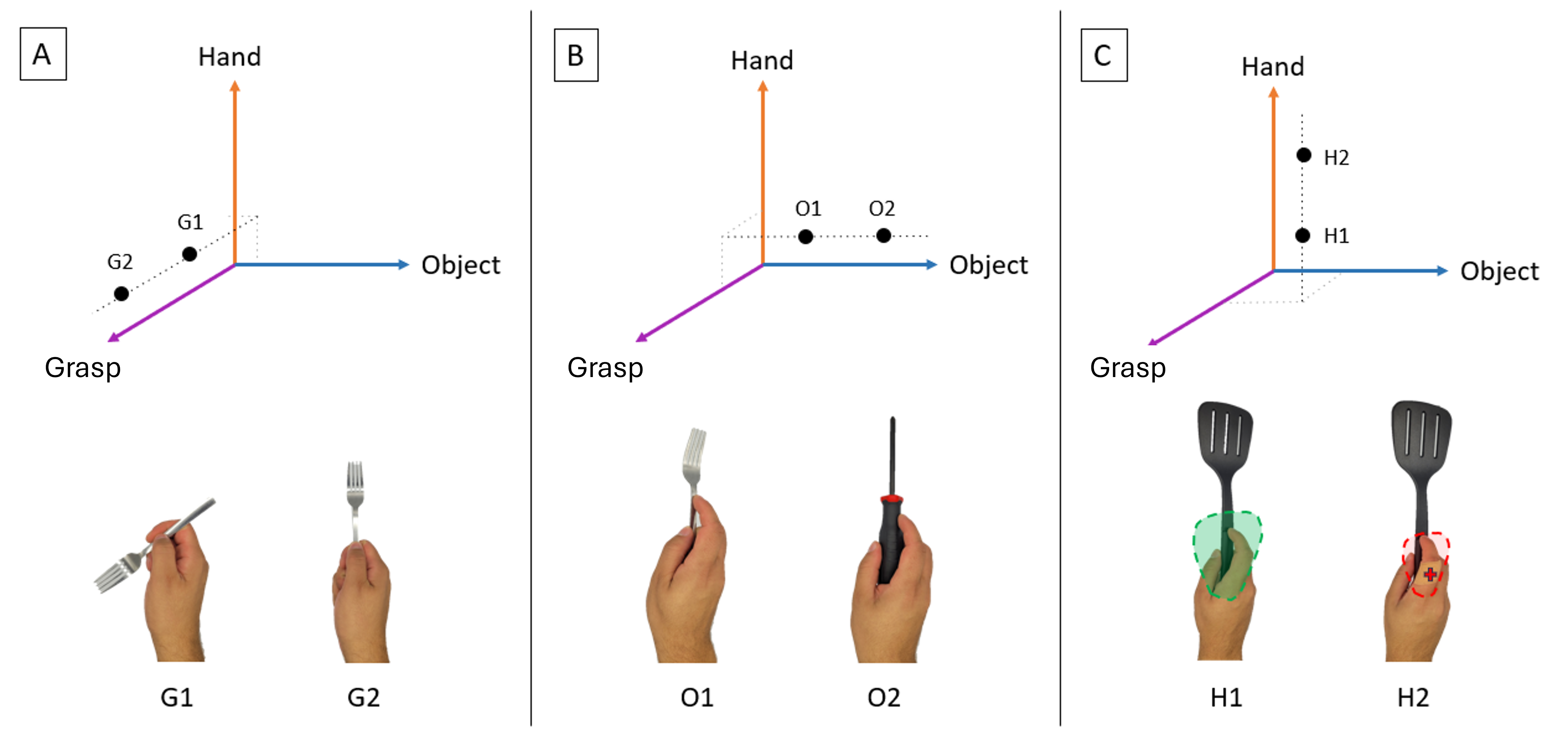}
    \caption{Design space for GPUIs. (A) represents a change in grasp type keeping the user's hand and object constant. (B) represents a change in an object while keeping the other two axes constant. (C) represents the same object and grasp type for the same user but in condition H2 the thumb has an injury that reduces its range of motion.}
    \Description{Design space for GPUIs. (A) represents a change in grasp type, keeping the user’s hand and object constant. The picture shows the 3 axes of the design space with a line representing change only in the Grip Axis and two example pictures of different grips with a fork (B) representing a change in an object while keeping the other two axes constant. The picture shows the 3 axes of the design space with a line representing change only in the Objects Axis and two example pictures of different objects, a fork and a screwdriver (C), representing the same object and grasp type for the same user but in condition H2 the thumb has an injury that reduces its range of motion. The picture shows the 3 axes of the design space with a line representing change only in the Hand Axis and two example pictures of hands holding a spatula, one with an injury in the thumb.}
    \label{fig:dspace}
\end{figure*}

\subsection{GPUI Design Space}
\label{sec:design_space}
Lastly, we construct a GPUI design space with three axes: grip, object, and hand (Figure~\ref{fig:dspace}). These fundamental variables always need to be considered when defining the hand-object coupled interaction for GPUI design.

\paragraph{\textbf{Grip Types}} Despite individual variations, most grips fall under one of the types presented in the grip taxonomy \cite{feix2015grasp}. The GPUI designer can begin by approximating the user's grip, using the taxonomy as a reference (Section~\ref{sec:grav}). Figure~\ref{fig:dspace} part A provides an example of this. 
    
\paragraph{\textbf{Grasped Objects}} Another axis is the grasped objects. Designers can traverse the space keeping the grip type and the hand as constants, and changing the grasped objects as seen in Figure~\ref{fig:dspace} part B. 
    
\paragraph{\textbf{Hand Type}} Hands are not the same. They vary in size, motion range, strength, and comfort preferences across users. Hands change over time, temporarily or long-term due to injury, disease, or fatigue. Variations can also be as simple as considering a right vs. a left hand. Figure~\ref{fig:dspace} part C shows an example where both the grasp type and the object are constant but because the thumb has an injury the thumb RoM is greatly reduced. 

The last factor that can be adjusted is the XR application. The interface layout needs to take into account the grasp, the user's hand, and the grasped object, but the designer can choose how to balance these elements. For example, they could keep the grasp and object constant while adjusting the application and related UI elements to design an ergonomic and comfortable user experience with UI elements that are always reachable regardless of their number or type.

\begin{figure}[!t]
    \centering
    \includegraphics[width=0.75\linewidth]{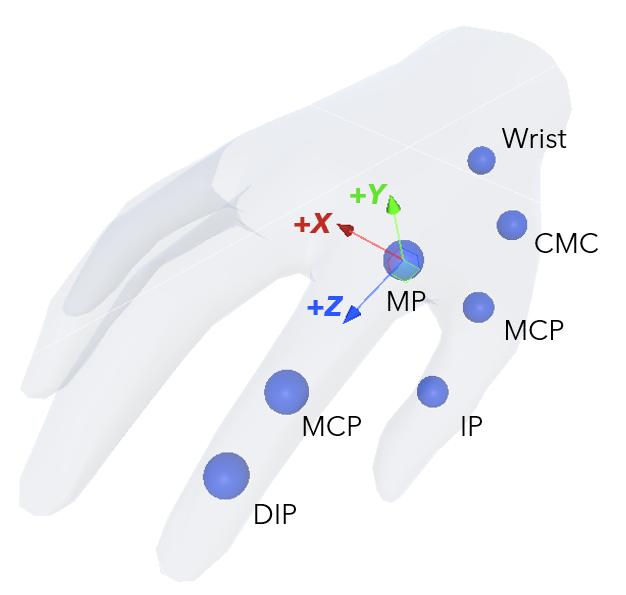}
    \caption{Procedural hand generated in the simulation process from anthropometric parameters. The forward direction points in the direction of the index. The right direction points from the index MP to the little finger MP. Positive rotations around the X-axis are counterclockwise.}
    \label{fig:hand}
    \Description{The procedural hand is generated in the simulation process from anthropometric parameters. Blue spheres represent the joints. The index is used as an example to show the corresponding joints from the wrist to the index's DIP joint. The forward direction points in the direction of the index. The right direction points from the index MP to the little finger MP. Positive rotations around the X-axis are counter-clockwise.}
\end{figure}

\section{Grasp Interaction Volume}\label{sec:grav}

RoM, or range of motion, is crucial for understanding the potential movement and reachability of the grasping hand. It is defined by the full extent of each finger's movements such as flexion, extension, adduction, abduction, pronation, and supination~\cite{chao1989biomech}. Since each joint, such as the wrist, knuckles, and fingers, has a limited range of movement, those limits form a boundary for the overall RoM. The motion boundary and positions of these joints relative to each other and to the grasped object form a complex set of constraints, which have implications for the design of ergonomic applications in various fields such as industrial design, robotics, and rehabilitation~\cite{lee2015ergonomic}. Techniques based on a joint's RoM also have the potential to assist designers of XR user interfaces in making data-driven layout decisions as demonstrated in XRgonomics \cite{evangelista2021xrgonomics} by the mid-air arm interaction space or by a motion range semi-sphere attached to the wrist~\cite{xu2018hand}.

\pname~enables access to the \textit{Reachable Interaction Volume (defined in Section~\ref{sec:riv})} spanned by the hand's RoM that can help designers create reachable and comfortable interfaces. The shape of the volume is restricted by the surface of the grasped object, self-collision of the fingers, and the limits of the hand RoM. In addition to the 3D volume, \pname~encodes the joint rotation cost with respect to an initial pose and is visualized as a color-coded map in the point cloud. We built a \pname~dataset and a simulation tool in Unity to make it easy for designers and developers to access and customize \pname's for their user scenarios. Besides using the volumes with the grasp types and objects we provide in the \pname~dataset, designers can simulate their own volumes by setting customized joint orientations and object surfaces. These input parameters can be collected through methods such hand tracking~\footnote{\url{https://developers.google.com/mediapipe}} and photogrammetry~\footnote{\url{https://poly.cam/}}, hand-object interaction datasets~\cite{hasson2019learning, chao2021dexycb,fan2023arctic,corona2020ycb_affordance}, or predicting hand grasps~\cite{omid2020grabnet, miller2004graspit} over a dataset of object 3D meshes~\cite{berk2015ycb}, among other approaches.

\subsection{Hand and Grasp Parameters} \label{sec:hand-params}

The anatomy of the hand is intricate and can differ between populations and change over an individual's lifetime due to factors like growth, injury, or disease. Our simulation technique takes a range of input parameters that designers can modify to represent a variety of hand types. Designers can specify the relative positions of joints in a hierarchy, to control essential anthropometric measurements such as hand length, breadth, and maximum spread. Furthermore, designers can define the RoM for each joint by setting minimum and maximum angles around each axis, making it possible to simulate interaction volumes for hands with limited motion or hyper-extension. Properly defined RoM values can prevent fingers from bending in unnatural ways. Hand thickness can also be parameterized along each inter-joint segment, which allows designers to simulate various levels of contact between the finger pads and the grasped object. While we provide a large set of interaction volumes in our \pname~dataset, designers can simulate their own volumes by customizing anthropometric parameters in our Unity simulation package, which we will make available along with our dataset upon publication 
(Section~\ref{sec:d-eval}). Designers can also choose to obtain joint hierarchy, positions, RoM, and hand thickness from external sources such as anthropometric databases~\cite{chao1989biomech}, or neural networks-based hand models~\cite{javier2017mano, yuwei2022nimble}.

Consistent rotation orientations are critical for accurate simulations. We keep the same 20 joint hierarchy as defined in the normative model of hand~\cite{chao1989biomech} while modifying the joint orientations to match Unity's left-handed coordinate system. The wrist is linked to the fingers along the joints MP (Metacarpophalangeal), PIP (Proximal Interphalangeal), and DIP (Distal Interphalangeal) in that order. The wrist is connected to the thumb along the joints CMC (Carpometacarpal), MCP (Metacarpophalangeal), and IP (Interphalangeal) in that order.

We adopt the following conventions for joint orientations, also depicted in Figure~\ref{fig:hand}. The forward direction (positive Z) of each joint is defined by the normalized vector pointing from the parent joint to the child joint. In the case of the wrist joint, the forward direction is determined by the normalized vector pointing from the wrist joint to the centroid of the hand. The right direction (positive X) of the fingers' joints is defined by the vector pointing from the little MCP to the index MCP. The right direction of the thumb joints is defined by the cross-product between the vector pointing from the CMC to the MP and the vector pointing from the MCP to the IP. We define these conventions to ensure a consistent reference for rotation directions across all the joints in our simulations. 

Since the hand cannot intersect with the object or with itself, the finger self-collisions, finger-palm collisions, and finger-object collisions
are discarded in the generation of the reachable interaction volume. To cover a wide variety of grasp configurations, our technique allows designers to input an object's surface geometry as a 3D mesh, an initial grasp pose defined by joint rotations, in addition to hand parameters as mentioned in Section~\ref{sec:hand-params}.

\subsection{GraVSim}\label{sec:sim}
Based on the design factors presented in Section~\ref{sec:factors}, we developed a tool in Unity that can simulate the motion of a grasping hand and generate GraV in the format of 3D point clouds color-coded by motion cost. To create GraV, GraVSim needs an object's surface, a joint hierarchy, and joint range of motion, initial rotations, and positions. GraVSim generates the \textit{Reachable Interaction Volume} by independently simulating the motion of each fingertip in the grasping hand. Selecting the fingertips as the \textit{Interaction Element}, the interaction volume of a hand includes all the points in the volume that can be reached by any movable fingertip, and conversely, any point within the interaction volume can be reached by those fingertips. Any positions where any part of the finger that intersects the object's surface are excluded from this volume.

\begin{figure}[!t]
    \centering
    \includegraphics[width=.65\linewidth]{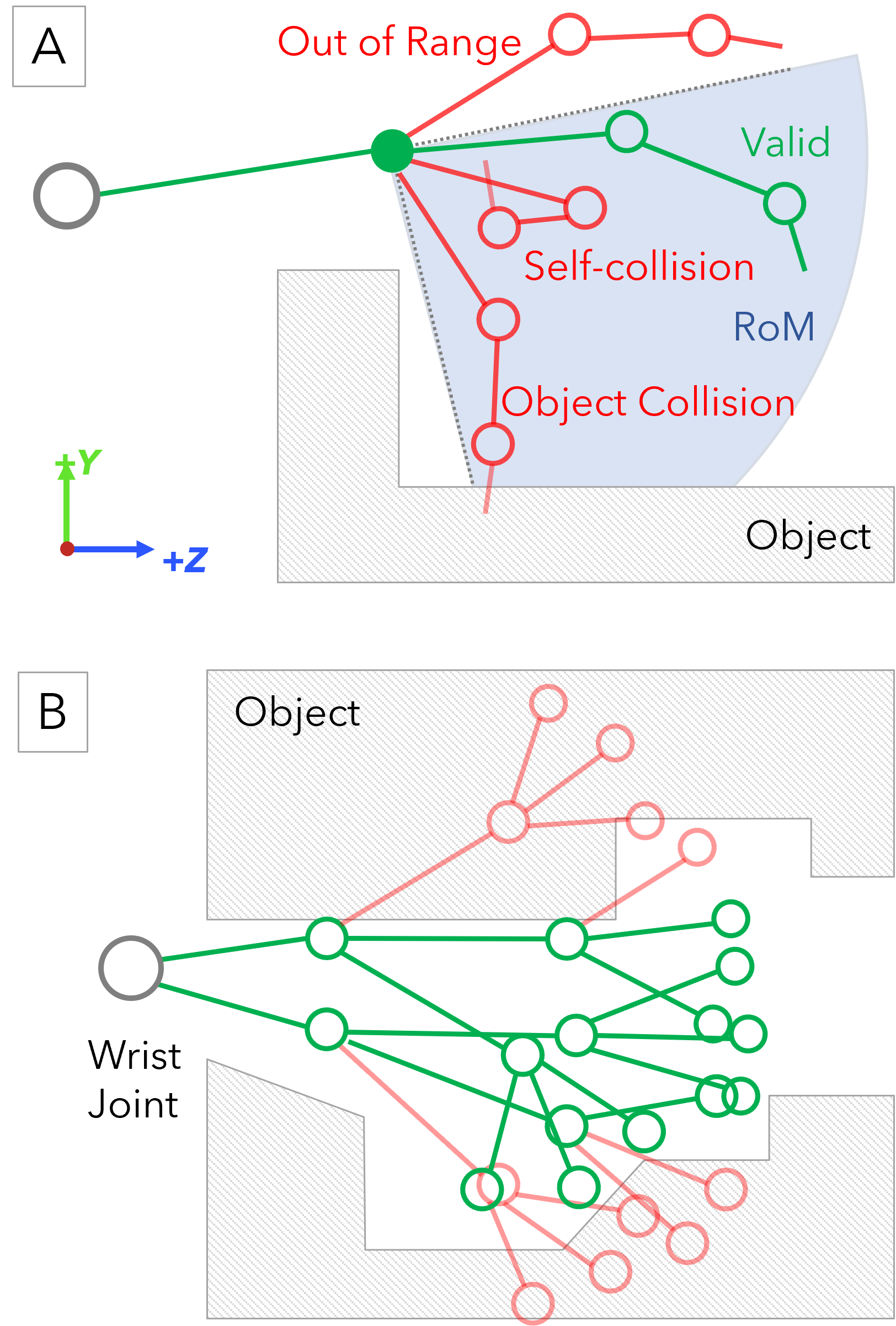}
    \caption{\pname~Simulation Process. (A) the simulation at the joint level stops when the finger either: goes beyond the limits of the RoM, collides against the hand, or collides with an object. (B) The execution expands along the joint hierarchy forming a tree-like structure.}
    \Description{GraV Simulation Process. Shows a vertical view of a hand armature. On the left, subfigure (A) overlaid armatures depicting the cases where the simulation stops. There is an L-shaped object underneath the armatures. There are three armatures in red: one that surpasses the line of the limit of the range of motion, another that touches the object, and a third one that curls in and touches the hand. Subfigure (B) depicts the execution trace of the simulation. It shows a series of overlaid hands. The hand is constrained all around by an object. The armatures in red intersect the object; the green ones do not.}
    \label{fig:tree}
\end{figure}

The simulation method employed by GraVSim is based on a Flood fill algorithm, which is behind paint bucket tools in many graphic editors. Flood fill finds connected regions adjacent to an initial pixel of an image limited by sharp boundaries~\cite{smith1979tint}. Similarly, we use the flood fill algorithm to explore the hand RoM by rotating a joint in discrete angular steps to iterate through adjacent rotations. In our case, the filling process stops at the boundaries of the RoM or when the finger collides with a grasped object or against the hand. Figure~\ref{fig:tree}a shows the stop conditions. Our simulation applies the flood fill algorithm for each finger independently starting from the joint closer to the fingertips (DIP of the fingers and IP of the thumb) until the root of the finger (MCP of the fingers and CMC of the thumb). After completing the flood fill in an initial joint, the simulation proceeds to the parent of that joint, taking a single filling step, i.e., just exploring the immediate neighborhood of that joint's current rotation. After this single step, the simulation executes an entire flood fill procedure for the child of that joint once again. This exhaustive hierarchical execution allows us to traverse the adjacent valid joint configurations with respect to the initial hand pose and find a connected volume where the fingertip can move unobstructed. We call this volume GraV. Figure~\ref{fig:tree}b shows a diagram of the execution tree process reaching valid joint configurations.

\begin{figure}[!t]
    \centering
    \includegraphics[width=.45\textwidth]{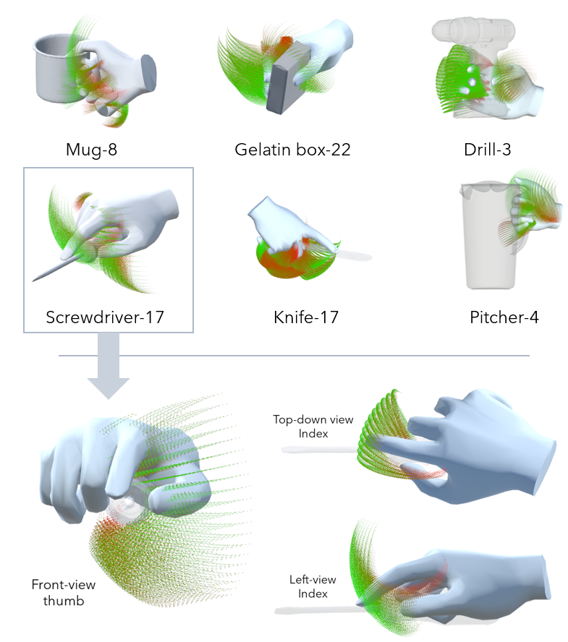}
    \caption{\pname~Dataset contains 367 grasp interaction volumes from feasible combinations of 33 grasp types and 58 objects available in YCB Affordance~\cite{corona2020ycb_affordance}. \pname~instances are labeled according to object type from the YCB object set~\cite{berk2015ycb} and a number from 1 to 33 indicating the grasp type according to the GRASP taxonomy~\cite{feix2015grasp}.}
    \Description{This figure presents a series of examples from the GraV Dataset. The dataset contains 367 grasp interaction volumes from feasible combinations of 33 grasp types and 58 objects available in YCB Affordance [ 6 ]. On the top part, this picture shows 6 instances of different objects and grip types with their generated volumes based on the YCB object set [ 3] and a number from 1 to 33 indicating the grasp type according to the GRASP taxonomy [15]. The bottom part of the picture expands on one of the instances, the screwdriver with grip 17, and shows this in greater detail from 3 angles.}
    \label{fig:dataset}
\end{figure}

At each step, the simulation records the position of the fingertip and a cost metric, resulting in a 4D point cloud. We adopted the total joint rotation cost, calculated as the sum of the joint angles between the current pose and the initial pose from all the finger joints. To manage computational complexity, we traverse the RoM of a joint in discrete angular steps defined by a minimum angular step parameter. Smaller angular steps result in denser volumes and longer execution times, while larger angular steps result in sparser volumes and shorter execution times. 

GraVSim is implemented in Unity, which allows designers to import and manipulate its input parameters directly in the same ecosystem as other XR SDKs they use for building XR experiences. Additionally, we offer a command-line interface that facilitates the manipulation of volume data and conversion into CSV, PLY, and OBJ formats for export into other tools.

\begin{figure*}[!t]
    \centering
    \includegraphics[width=0.9\textwidth]{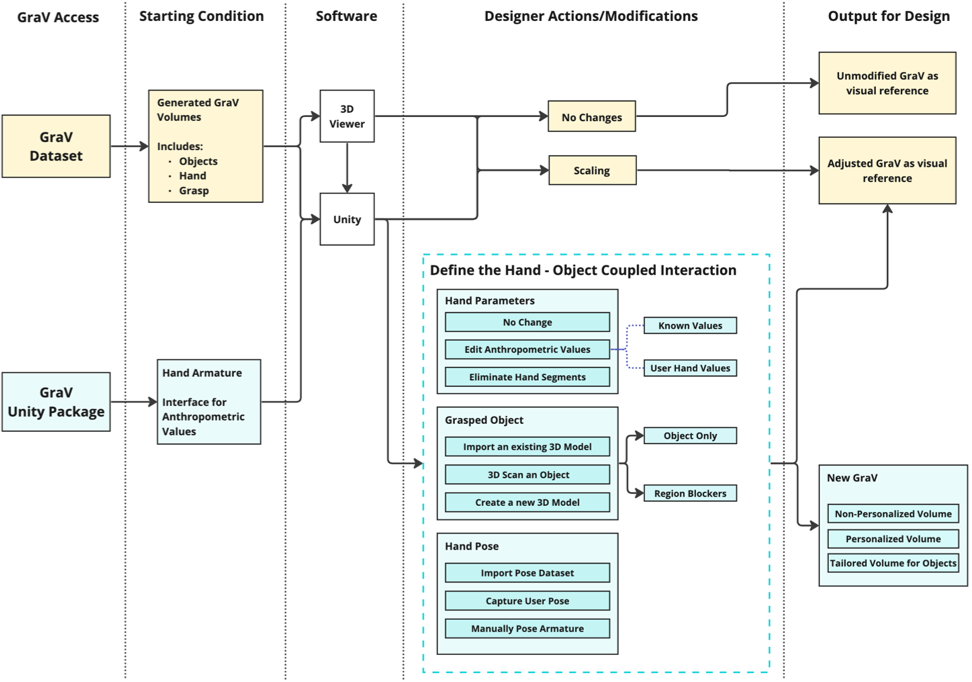}
    \caption{Two different pathways for designers to access GraVs: pre-created GraVs from our dataset (yellow) and the Unity Simulation Tool to create new GraVs (blue). }
    \label{fig:workflow}
    \Description{Representation of the pathways designers can take to use the dataset and the simulation tool. The dataset can be used as is, or it can be scaled, while the simulation tool requires the hand pose, hand parameters, and grasped object to generate a new GraV volume. It also shows where in the workflow designers can add Region Blockers to address parts of the objects or the hand’s vicinity they don’t want users to reach.}
\end{figure*}

\subsection{GraV Dataset}
\label{sec:grav_dataset}

To support designers creating GPUIs and demonstrate the feasibility of our simulation procedure, we generated a \pname~dataset containing 367 grasp interaction volumes. We build our dataset using as input parameters grasp poses and objects from the YCB Affordance dataset~\cite{corona2020ycb_affordance}, which is a publicly available dataset that contains each of the 33 grasp types proposed in the GRASP taxonomy~\cite{feix2015grasp} manually annotated over 58 objects from the YCB object set~\cite{berk2015ycb}. We provide \pname's in the form of CSV, PLY, and OBJ files labeled with object and grasp types. The minimum angular step in the simulations to generate the \pname~dataset was 5 degrees. Figure~\ref{fig:dataset} shows examples of grasp interaction volumes that we generated. The annotated grasp poses in the YCB Affordance dataset are provided as MANO representations~\cite{javier2017mano}, a neural network-based hand model capable of representing a wide range of hand types and poses. We access the joints provided by the MANO representation to initialize the joint positions in our simulations to create the interaction volumes. The joint hierarchy is defined according to the normative model of the human hand~\cite{chao1989biomech}. 
We calculate the joint orientations based on the joint positions to match the conventions presented in Section~\ref{sec:sim}. We initialize the hand joint RoM in our simulations with the values provided in the anthropometric dataset~\cite{chao1989biomech}. We initialize the thickness of the hand in our simulation with the median hand thickness of the male American population~\cite{chao1989biomech}. To reduce parameter complexity, hand thickness is assumed to be constant over the entire hand.

\subsection{Designer Workflows}
As shown in Figure~\ref{fig:workflow}, \pname~dataset and the Unity Simulation package can both be used (and modified) by the designer in a variety of ways to provide volume data to inform UI placement. Designers can import the dataset or individual volumes from the dataset into Unity or other compatible software to access the 3D representation of the hand-object coupled interaction and the reachable volume. They can use the volume data as-is to assess reachability or scale it to explore UI designs for different hand sizes. The Unity Simulation package includes a poseable hand armature and an interface to modify hand anthropometric parameters. In this workflow, designers can freely define their own hand parameters, import a 3D object model, and choose a grasp that best represents their user scenario. In addition, they can designate certain areas on the grasped object or surrounding space as Region Blockers. These blockers aim to guide designs toward specific reachable regions, to either constrain movement or prevent access to hazardous parts of the object (such as hot regions or blades) or places already equipped with physical interfaces (like buttons, switches, or sensors).

\section{Design Workshop}
\label{sec:d-eval}
\label{sec:eval}
To assess \pname's applicability, we conducted a design workshop, individual questionnaires, and group discussions with XR designers, developers, and researchers. It was approved by our local IRB (protocol 7-22-0512) and participants provided informed consent. Our study aimed to: 
\begin{itemize}
    \item Explore how XR developers could approach the design of grasp-proximate user interfaces supporting physical tasks.   
    \item Understand how GraV data (reachability and displacement cost) can inform design decisions, how designers would incorporate it in their workflow, and recommendations to improve GraVSim. 
\end{itemize}

The design constraints were derived from presented design factors (Section~\ref{sec:factors}) while the explored cases traversed the axes of the design space. To prevent any biases during the exploration of the designers, the design factors (Section~\ref{sec:factors}) were omitted from the workshop, and only the design space (Section~\ref{sec:design_space}) was mentioned to introduce the design scenarios. 

\subsection{Participants}
We recruited nine participants (5 males and 4 females; age range: 20-34). All participants were familiar with XR and UI design as they were either professional software developers or VR/MR researchers with 1-5 years of VR/MR experience. All participants mentioned Unity as one of their primary design tools. Some of the other tools they had experience using were Unreal Engine, Snapchat Lens Studio, Facebook Spark, AR Core, and AR Kit. Previous work included the design of co-located social games in AR, interfaces for augmenting spatial-awareness MR, real-time MR task guidance system design, open space navigation with AR assistance, location-based immersive performance design, and interfaces to facilitate interaction between handicrafts and digital fabrication. 

\begin{figure*}[!t]
    \includegraphics[width=1\textwidth]{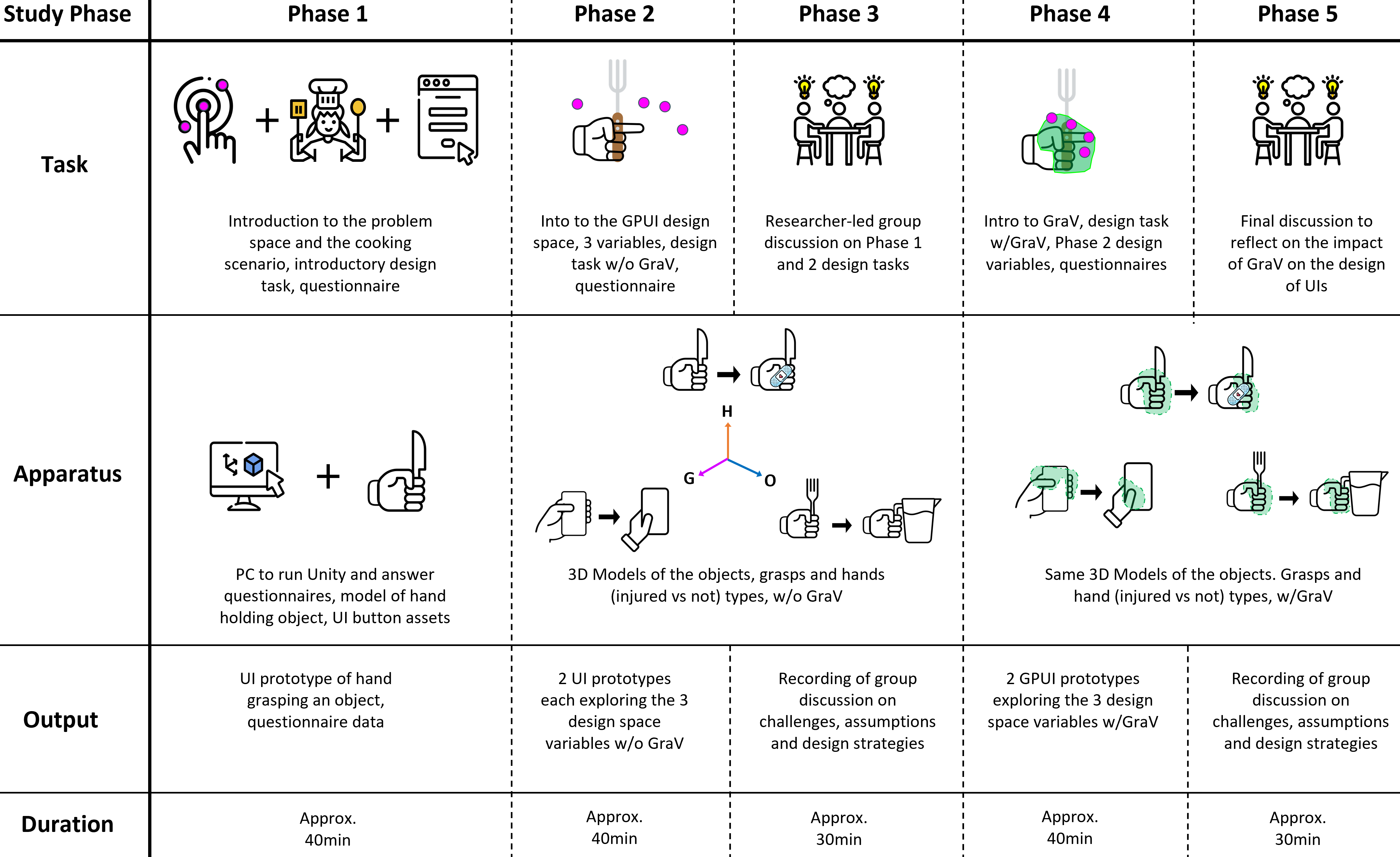}
    \caption{Overview of the GraV formative study procedure consisting of three design tasks (Phases 1, 2, and 4) and two group discussions (Phases 3 and 5), where XR designers created 3D UIs with- and without-\pname data.}
    \label{fig:grav-study-design}
    \Description{The figure shows a table with 5 rows. The top row is the phases, the second row is the tasks, the 3rd row is the apparatus used in the study, the 4th is the output, and the last row is the duration of each task. The figure shows each of the phases of the user study. Phase Columns 1, 2, and 5 show icons representing the design tasks and briefly describing each. Phase 1 and 2 were design tasks without GraV. Phases 3 and 5 columns present a brief overview of the discussion points covered during the group interview.}
    \Description{}
\end{figure*}

\subsection{Apparatus}
The user study was conducted in two group sessions. Due to geographical location and time zone differences, we opted for a hybrid approach combining in-person and video conferencing using Zoom. To support seamless communication between the remote and local participants, we projected the virtual participants onto a large screen. We used the Meeting OWL\footnote{\url{https://owllabs.com/products/meeting-owl-3}} three with a 360-degree camera, speakers, and microphone capable of bringing into focus one or more speaking participants. For the study design tasks, participants used Unity as their design environment and were provided study assets to download at the start of the study. The 3D assets included the objects, hand model with the selected grasp, and generated reachability point clouds with color-coded displacement costs (GraVs). \change{We provided physical props for the designers who preferred to interact with the physical objects while designing. For the remote participants, a list of the objects was sent before the workshop with minimal specifications to prevent bias. Since the items were common household kitchen items, they were instructed to acquire equivalent items in case they could not find exact replicas. Participants were informed during the workshop that using these objects was optional, allowing them to reference them based on their preferences.}

\subsection{Evaluation Design and Procedure}

\subsubsection{Workshop Design}
We conducted two 3-hour workshop sessions, with 5 and 4 participants each. Our study consisted of a study setup and 5 phases as summarized in~\ref{fig:grav-study-design}. To facilitate the exploration of the GPUI design process, our study consisted of three UI design tasks. An introductory design task to introduce the design scenario and constraints of one-handed UI design for the physical task (Phase 1), a design task without GraV data (Phase 2) to explore the GPUI design space (presented ~\ref{sec:factors}), and finally with GraV data we explore how the provided data could be used by designers in approaching the design of GPUIs. For the \textit{With-GraV} and \textit{Without GraV} design phases, to provide a set of constraints that also traversed the three axes of the design space presented in Section~\ref{sec:design_space}, we randomly split the participants into three groups. Participants in these groups were assigned different interface design conditions where either the grasp, the object, or the hand type changed (e.g., injured finger). For the UI prototype, participants focused on designing a simple, widget-based UI that could be used to interact with an AR task guidance system for cooking. The intention of the design tasks was not to assess the appropriateness of their UI designs but to use the design tasks as constrained scenarios for the participants to explore the intricacies of creating GPUIs. 

The workshop's intent was not to evaluate a system but to explore how designers could approach the GPUI design space and use the GraV data in their design workflows while maintaining a certain level of agency. \change{To capitalize on the participant's initial impressions of GraV and their preliminary approach to incorporating reachability information into their workflow, we decided to focus on a short exploration. To maintain a minimum iterative design workflow, the workshop included multiple design tasks that built upon previous tasks, creating a continuous effort throughout the study.} Participants transitioned from the baseline (Without-GraV) condition to the alternative condition (With-GraV), allowing for a direct comparison of their experiences in designing UIs using both processes. The decision to maintain a consistent order of conditions is based on considerations aimed at planning for ecological validity~\cite{hartson2012503}. This approach aligns with real-world scenarios where individuals would be familiar with a design process, similar to the baseline, before engaging with GraV. Additionally, we identify the potential for asymmetric skill transfer between conditions, where exposure to reach and cost information from GraV may influence behavior in the baseline condition, while the baseline does not yield new insights for the alternative~\cite{mackenzie2013157}. Lastly, we anticipated minimal additional learning about Unity, the underlying platform of the study, given participants' extensive prior experience with the tool. 

\subsubsection{Design Scenario Context}
To provide context about the use of the objects and the target user interactions, the provided scenario was a novice chef using an AR task guidance application that provides them with instructions on how to accomplish a cooking task. The chef needs to be alert at all times and keep their eyes on the cooking task while interacting with the interface (without putting down the tool) to follow the instructions, hence the need for GPUIs. All design tasks consisted of prototyping a UI for the provided cooking scenario. To simplify the task, all participants were provided with 4 UI widgets for navigating forward and back through task instructions, viewing the full recipe, and watching a video for help with any step. These widgets were used in all three design tasks.

\subsubsection{Data Collection and Analysis}

We collected individual questionnaires and conducted researcher-led group discussions. The combination of the design tasks and both individual forms and group discussion questions were chosen to help elucidate not only the challenges that participants encountered but also their underlying assumptions and the design strategies they employed to overcome these challenges. Participants uploaded their UI designs at the end of each phase. For the questionnaire, open-ended questions focused on their thought process, design justifications, and any insights. For the discussions, a researcher guided the conversation with open-ended questions, and all participants were encouraged to respond and build upon each other’s comments by expressing similar or contrasting experiences during their design tasks. The focus of the group discussion was on the challenges, assumptions, strategies, ergonomic assessments, and potential uses of GraV (Phase 4) they identified during the study. We analyzed both the prototypes and feedback in a two-step process. First, two researchers analyzed the prototypes independently to identify patterns. Following that, the researchers analyzed the feedback to categorize themes correlated with the UI design prototypes, resulting in six main clusters. Section~\ref{sec:d-results} provides more details about the resulting clusters and insights from the designers.

\subsection{Design Tasks} \label{sec:d-eval-tasks}

The design tasks were presented during Phases 1, 2, and 4 of the study (Figure~\ref{fig:grav-study-design}). In this section, we present the design constraints for each of the design tasks as well as the objects, grasp types (based on the GRASP taxonomy~\cite{feix2015grasp}), and hand types.

\subsubsection{Phase 1: Introductory Design Task}
For this task, participants were provided with a 3D model of a hand grasping a knife Figure~\ref{fig:design_intro}. This task served as an initial exploration of the constraints related to GPUI design and as a clarification exercise on doing the design task and uploading their build and design documentation.

Participants were provided with the following specific design constraints:
\begin{itemize}
    \item The UI should be reachable by the fingers holding the object, without dropping it or requiring the other hand. 
    \item No gaze, gesture, or voice commands.
    \item The interface can only be created with the provided UI widgets.
    \item Each UI widget must be reachable by at least one finger, but multiple-finger access is acceptable. 
    
\end{itemize}

\subsubsection{Phase 2: Without-GraV Design Task}
For this task, the axes of the design space (grasp type, object, and hand type) were introduced. Participants were randomly split into three groups and each group was provided with two 3D models representing their group's respective axis. The object group had two different objects (a water pitcher, and a fork) with the same grasp and hand type. The grasp group had two different grasps (grasp 22 and grasp 31 from the taxonomy \cite{feix2015grasp}) with the same hand and object. The hand group had two different hand types (injured finger, uninjured hand) with the same grasp and object types. For the hand condition, the model was a hand holding a knife using a power grip, half of the index finger was removed from the second model of the hand to represent an injury. 

The design constraints for this task remained the same as the introductory design task with the following additions:

\begin{itemize}
    \item The user can reach the UI elements at all times without extending their arm. 
    \item One prototype for each assigned 3D model representing a change in grip type, object, or hand type. 
\end{itemize}

\subsubsection{Phase 4: With-GraV Design Task}
For this task, we introduced the participants to GraV and explained how the point cloud was generated and the cost metric calculated to colorize the point cloud. To allow for a more open exploration participants were told to use their own discretion and use the information provided by GraV, based on their design criteria, instead of simply placing the UI widgets within the volume. 

The design constraints for this task also remained the same as the introductory design and phase 2 tasks with the addition of the use of GraV along with the same 3D models provided for the Phase 2 baseline design task. 

\section{Results} \label{sec:d-results}

\subsection{Interfaces Designed by Participants}
Each of the 9 designers in the workshop produced one UI for the introductory task and two UIs each in Phases 2 and 4, for a total of 36 interfaces and 144 button placements. This section provides an overview of the evaluation of their UIs and results obtained from these qualitative inquiries in the workshop.

\begin{figure}[!t]
    \centering
    \includegraphics[width=\linewidth]{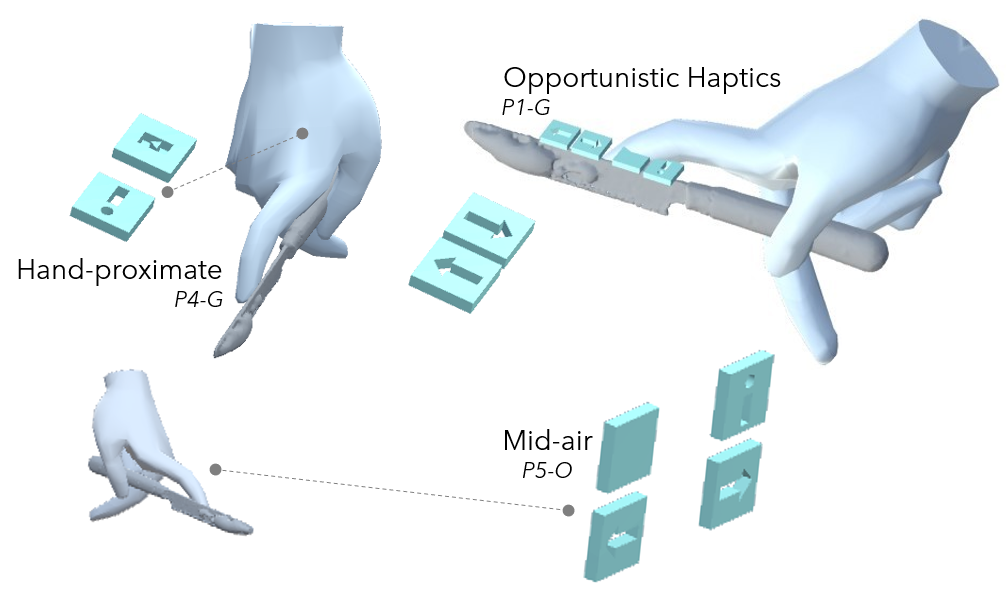}
    \caption{Designs created by participants in the introductory task for point grip knife blade 3D model from the YCB dataset \cite{berk2015ycb}, showcasing each of the three identified categories. The ``Opportunistic Haptics'' prototypes aligned buttons with the knife's surface, enabling passive opportunistic haptic feedback (P1). "Hand-proximate" prototypes featured buttons intended to be reached by the fingers (P4). In the "Mid-air" prototypes buttons were accessible by arm motion and outside finger reach (P5).} 
    \Description{Designs created by participants in the introductory task showcasing each of the three identified categories. The “Opportunistic Haptics” prototypes aligned buttons with the knife’s surface, enabling passive opportunistic haptic feedback (P1). "Hand-proximate" prototypes featured buttons surrounding the hand intended to be reached by the fingers (P4). In the "Mid-air" prototypes, buttons were accessible by arm motion and with the outside finger's reach (P5). This picture shows a hand holding a knife with a floating interface in front of the user’s hand.}
    \label{fig:design_intro}
\end{figure}

\paragraph{Introductory Designs} The UI designs for the introductory task (Phase 1) were clustered into three groups: (1) opportunistic haptics - designs where UI elements are aligned with the held object's surface enabling passive haptic feedback; (2) hand-proximate - designs where UI elements are intended to be reachable by the fingers; and (3) mid-air - designs where UI elements are not within finger reach and would require arm motion to be reached. Figure~\ref{fig:design_intro} shows examples from each category. The point grip on the knife blade (often used by sushi chefs\footnote{\url{https://web-japan.org/trends/11_tech-life/tec202211_japanese-knives.html}}), was taken advantage of by designer P1-G (based on their blade handling experience) to place all UI buttons along the dull edge of the blade(Figure \ref{fig:design_intro}). 

\paragraph{Design Variations}
We clustered the designs from Phases 2 and 4 into three groups, presented in Figure~\ref{fig:design_change}. The clustering is based on how the designs changed from those created in Phase 2, before using \pname, to designs created in Phase 4, after using \pname. The three categories are, (1) Layout Change - groups together the designs with changes in the UI element's position, rotation, or scale to fit the \pname; (2) Interaction Change - groups the designs that changed to allow interaction with fingers and motions not previously considered; and (3) Goal Change - groups the designs that changed to support a design goal such as reducing accidental interactions referred to as the Midas Touch problem \cite{istance2008snap} or favoring interactions with functionalities of assumed priority.

\subsection{Feedback from Designers}
Through the individual questionnaires and the group discussions, we aimed to learn about participants' prior knowledge and design processes and how they might apply those to one-handed UI design, the expected challenges for one-handed interactions, and potential design strategies to overcome the GPUI constraints. After presenting everyday scenarios, in the introduction design task, such as using a screwdriver or drill to assemble furniture, participants expressed that they could relate to the disruption flow and fatigue when interacting with mid-air interfaces. They also acknowledge the challenge of one-handed interaction design, particularly when compared with more familiar approaches like mid-air, on-body, and on-object interactions requiring two hands. Furthermore, none had previously addressed the grasp-proximate one-handed scenario we presented, and they were not aware of existing tools, data, or strategies that could help with the design of GPUIs. These insights reinforce the key challenges in designing XR for physical interactions, as previously presented by Ashtari et al \cite{ashtari2020creating}.

\subsubsection{Without-GraV} Below, we categorize participants' input into three categories, derived from the initial discussion (phase 3) and individual questionnaires, centered around tasks formulated without GraV.

\begin{figure}[!t]
    \centering
    \includegraphics[width=\linewidth]{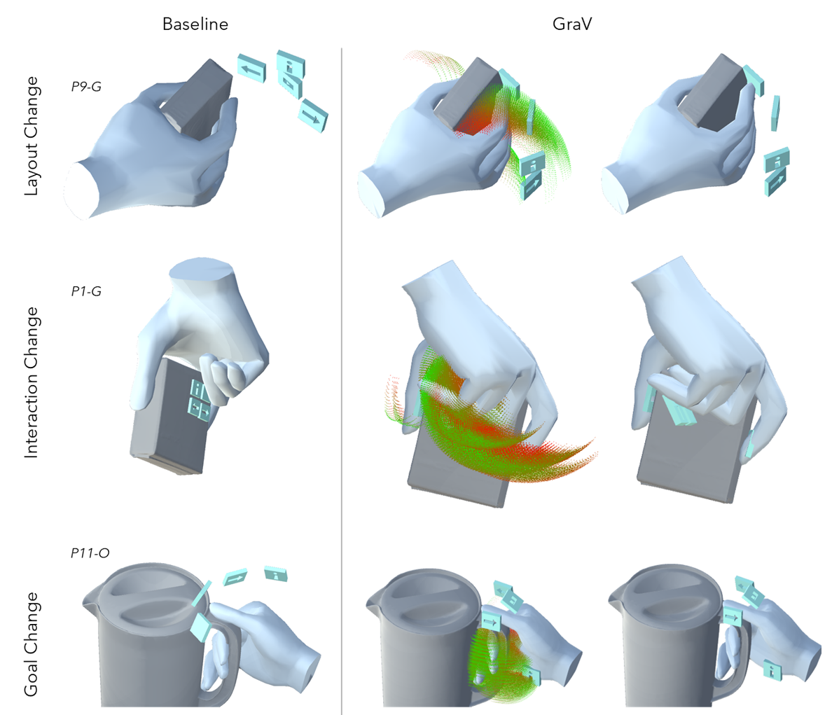}
    \caption{Designs created by participants using the three variables of hand, object, and grasp type in Phase 4 with \pname. In the Layout Change row, buttons in front of the box were moved to fit within the interaction volume (P9, grasp group). In the Interaction Change row, buttons were moved from the surface of the box to the area underneath the palm (P1, grasp group). In the ``Goal Change'' row, a UI widget was placed near the wrist to avoid accidental interactions (P11, object group). The leftmost column shows the designs Without-\pname~and the two columns on the right show the updated versions of the designs in the left column created With-\pname.}
    \Description{Designs created by participants using the three variables of hand, object, and grasp type in Phase 4 with GraV. There are three rows in this picture representing the major changes participants made to their layouts. In the Layout Change row, buttons in front of the box were moved to fit within the interaction volume (P9, grasp group). In the Interaction Change row, buttons were moved from the surface of the box to the area underneath the palm (P1, grasp group). A UI widget was placed near the wrist in the Goal Change row to avoid accidental interactions (P11, object group). The leftmost column shows the designs Without-GraV and the two columns on the right show the updated versions of the designs in the left column created With-GraV.}
    \label{fig:design_change}
\end{figure}

 \paragraph{Expressed Challenges} Participants expressed that the limited space available for the interface layout around the grasping hand was a challenge. Among the factors that limited the layout space, they identified the area the user could reach with their fingers, the surface of the object covered by the grasping hand, the grasped object itself and anticipated surrounding obstructions. Participants manifested concerns about obstruction between virtual and physical elements, e.g., interface elements occluding the field of view needed for performing the physical task, and physical elements blocking the interaction with interface elements. \textit{``Different environment configurations would impose different ergonomic constraints (e.g., something is blocking the buttons in front of you).''} - P5. Another commonly expressed concern was that reflex (mirror) hand movement could cause accidental interactions with the interface elements and result in undesired physical consequences, such as spilling the contents of a pitcher as the user reaches forward with the thumb~\cite{monteiro2021hands, kim2022pseudo}. P3 expressed this challenge by saying, \textit{``It is important to know when you are moving a finger...how that impacts other fingers' movement. Because if I am moving my ring finger...this one moves automatically (showing the little finger) so I could misclick a button.''} 

\paragraph{Adopted Task and User Assumptions} Participants presented a series of adopted assumptions during the design process. Some revealed that they assumed the user’s hand had a certain size and scaled the interface elements accordingly. Some assumed that certain fingers would always be grasping the object and therefore designed primarily for the index finger and the thumb, which they felt more familiar with. Participants also shared assumptions they adopted related to the priority of elements in the interface and functionalities of the application. 

\paragraph{Implemented Design Strategies} Participants communicated strategies they employed for designing grasp-proximate interfaces without the support of GraV. For instance, some grasped a real object that matched the object in the design task and simulated interacting with the UI while grasping that object to understand the interaction intricacies. P1 - \textit{``I am not sure how precise users could be. I needed to try to hold the object myself because otherwise, I would not have known how to hold it.''} Others said they simply imagined how far a user’s finger would reach and used that as a strategy to help them place the UI widgets. Participants restated the assumption about interface element priority and described the approach of placing high-priority elements next to fingers assumed to be dexterous to support easy access to high-priority features, but other participants mentioned they decided to place assumed high-priority elements closer to fingers that had less assumed influence on the grasp. P3 - \textit{``I tried to ensure the grip wouldn't be impacted by clicking buttons.''}

\subsubsection{With-GraV} Here we present the gathered information during the second discussion session and individual questionnaires based on the With-GraV design task. Figure~\ref{fig:design_change} shows the changes in participant designs using \pname~ from designs without-\pname. 

 \paragraph{Updated Assumptions} Participants declared that the grasp interaction volume replaced their finger reach assumptions. P2 wrote: \textit{``(GraV) helped counteract a tendency to overestimate the flexibility of user hand and finger movements.''} Participants voiced that after having access to GraV they rescaled the interface elements to fit within its boundaries. They stated that GraV provided data about which fingers were able to move by how much when grasping an object.

\paragraph{Revised Strategies} Some participants stated that they used GraV to refine the element positioning of their initial layout. P9 wrote, \textit{``I could determine the button positions based on the visualized accessible area.''} They mentioned favoring low-cost regions of the volume as a general strategy. Some participants expressed they placed elements of assumed high importance in the low-cost regions of volume. P4 -\textit{``Knowing the low and high-cost zones allowed me to place the buttons based on their importance/priority/frequency of being pressed.''} Others communicated they placed high-importance elements in the high-cost regions of the volume to avoid accidental interactions. \textit{``High-cost zones are less prone to false positives and allow smaller buttons'' - P9.} 

\paragraph{Expressed Benefits} Participants pointed out that GraV helped them better understand hand types that were different from their own. When asked how GraV would change their design process, P1 answered: \textit{``I would maybe try to use less of my own assumptions about body movements when for example designing for a child’s hand or a person with arthritis.''} Participants noted that they were able to make faster layout decisions with GraV. They also described how GraV helped them to identify design opportunities such as using fingers to interact in directions not originally considered such as moving outward and using the fingernails as the selected input part. They also said that it was helpful to consider other fingers, besides the thumb and index, for interaction. P1's design presents an example of this in the \textit{Interaction Change} row in Figure~\ref{fig:design_change}.

\section{Discussion}
In this section, we expand on the outcomes of the Design Workshop~(Section~\ref{sec:d-eval}). Overall, the resulting opinions of the designers resonate with the design factors and the intended use cases for GraV. Designers expressed that GraV provided valuable information that could assist them in better understanding the hand-object interaction and depend less on assumptions. 

\subsection{Elementary Approaches to GPUI design} 
Throughout the introductory design task, we noted that the assumptions and strategies employed by participants were based on their levels of confidence and prior experience with designs akin to XR interfaces for physical tasks while grasping an object.

Initially, designers (P2, P5, P6, P7) expressed a lack of confidence or familiarity with such interfaces. They tended to employ a familiar mid-air approach, while still ensuring they could maintain a grasp on the object. In contrast, participants (P1, P3, P4, P8, P9) who articulated concerns regarding interface visibility during object usage or exhibited some degree of familiarity with the design scenario opted for a more hand-proximate design approach. This approach is exemplified in Figure~\ref{fig:design_intro} in the P4-G example. Furthermore, it is noteworthy that two designers with backgrounds in digital fabrication or manual crafting incorporated the object surface to provide haptic feedback when interacting with the UI widgets. These findings provide insights into the challenges experienced designers may encounter when attempting to diverge from established spatial design practices, inspired by 2D UI design \cite{evangelista2021xrgonomics}, particularly when engaging with designs tailored for single-handed interactions. They also show how real-world familiarity with hand-object interactions could help designers feel more comfortable with GPUI design.

We observed that the challenges expressed by the designers in both the introductory task (Phase 1) and the without-GraV design task (Phase 2) reflected the factors presented in Section~\ref{sec:factors}. When we asked participants what information they used to complete their designs, they mentioned that knowing how the user could reach the interface was crucial. P8 said, \textit{``I tried to understand which fingers have more flexibility to move and what is the range of movement available''} - P8 continued, \textit{``I feel like having the haptic of the knife -oh I touched the knife-... feels potentially important.''} P5 shared the following, \textit{``I think the biggest challenge was thinking about how close or far the buttons should be from the hand since I had to balance between accidentally clicking the buttons and keeping them close enough to hit.''} This shows considerations that correlate with the Finger Motion Range and the Reachable Interaction Volume factors presented in Section~\ref{sec:factors}, before participants were introduced to GraV. 
P1 expressed concerns about the lack of clarity on the possible movements and how holding a physical object was useful to get a more tangible idea, saying, \textit{``Figuring out possible outcomes of movement, to me it was extremely helpful to hold the objects and I cannot imagine what it would be like to try to design without holding the actual object... it's hard to think of a UI that would actually work.''} We believe that for cases like this, the boundary conditions and reachability values can offer a clearer perspective on the limitations for both the object and the reach of the fingers.

\subsection{Towards GraV Adoption}

Designers shared how they would use it in their personal design workflows and future applications including expressed interest in a model that could incorporate more information about the biomechanics of the hand, additional visual elements (e.g., animated visuals of the hand behavior reaching certain areas of the space), and even the level of visibility of certain UI elements.

Some of the designers highlighted that GraV inspired them to design interactions for the back of the fingertip (specifically the fingernail). This departure from commonly used interaction modalities for XR interfaces was particularly notable. Some designers integrated this strategy into their UI designs as presented by Figure~\ref{fig:design_change} in the Goal Change row with pitcher design, where the designer places a button meant to be reached by the back of the thumb when using GraV.

Some designers highlighted the fact that the interaction costs visualized in the GraV had the potential to streamline certain parts of the design process. P1 said, \textit{``If I am going to see a point cloud I might as well just send that to an algorithm and ask it to optimize (...)''}, suggesting they foresee that GraVs could become input for some kind of UI layout optimization algorithm. Expanding on this possibility we envision GraV becoming a common exchange format, visually meaningful for humans and machine-readable, fostering human-AI design collaboration. We also believe this could facilitate collaboration across designer teams developing interfaces for various applications under the same hand-object coupled interaction, by providing a quantifiable common ground about their target user scenario. Designers commented on wanting to integrate this with other interaction modalities. P4 said, \textit{`` This combined with gestures could be a pretty powerful way to reduce the fatigue caused by multi-limb movement in AR and VR.''}

\section{Limitations \& Future Work}

Currently, we simulate finger motions independently from each other, i.e., a finger does not interfere with other fingers during the simulation, they only collide with themselves, the palm, or the handheld object. Given this limitation, our simulation cannot generate GraVs of interactions that require coordinated motion of multiple fingers, e.g., finger-thumb gestures. To accomplish that, it would be necessary to revise the definition of a valid fingertip position to consider the joint orientations of other fingers. Additionally, the simulation does not generate GraVs that represent involuntary co-dependent movement. Fingers are not completely independent from each other and present mirror movements, which are simultaneous involuntary movements in response to the voluntary movement of a different finger~\cite{nadkarni2012mirror}. This phenomenon has implications for GPUI design because it can cause accidental interactions, known as the Midas Touch problem~\cite{istance2008snap}. \change{Future work can consider biomechanically accurate simulations using musculoskeletal models, such as OpenSim~\cite{delp2007opensim}, to simulate more authentic hand-object interactions. This approach can inform designers of finger co-dependent movement and generate new types of finger motion costs.}

Another limitation of our current method for generating GraV is that it does not simulate Physics and focuses only on collisions. There are several variables to be considered for a physically accurate simulation including weight distribution along the object, static friction between hand and object, static equilibrium, and skin and muscle deformation along with the held object parameters. The designers recognized that the reachability was crucial information that is not evident in contrast to visible object elements (e.g., dangerous areas, existing physical buttons, moving parts, etc.).

\change{As presented in Section~\ref{sec:d-results}, designers found that GraV enhanced their awareness of the reachable space of a grasping hand. Although reachability is crucial for designing GPUIs, it is not the only factor. For example, keyboard designs often trade some reachability for better visibility. Future research should explore factors beyond those mentioned in this paper. A promising direction is the study of transitions between grasp types, which occur when users accommodate objects in their hands. Our design workshop considered two grasp types for the same hand-object pairing, but further studies on grasp transitions could enhance GPUI design process.}

\section{Conclusion}
In this paper, we introduce the motivation for how XR interfaces for physical tasks could benefit from adopting a similar design approach to tools and other everyday objects with integrated interfaces. Based on that, we present the idea of Grasp-Proximate User Interfaces (GPUIs), a series of influencing factors, and a design space to assist designers in approaching these types of UIs.

We also present Grasp Interaction Volume (GraV) for UI designers, which represents the 3D space that can be reached by the hand or fingers while holding an object, presented as a point cloud that encodes the joint rotation cost at each point. To provide access to GraVs, we built a dataset of GraVs, based on anthropometric values and common everyday objects, and a finger motion simulation tool in Unity (GraVSim) to generate new GraVs. Results from a workshop design evaluation of GraV with XR designers indicate the benefits and value provided by the interaction volume data for supporting the design of GPUIs. 

\begin{acks}
This work was supported by the U.S. National Science Foundation (Early CAREER Award 2023 no. 2240133). We thank Ana Cardenas, You-Jim Kim, and Ashley Del Valle for their helpful feedback and discussions.    
\end{acks}

\balance
\bibliographystyle{ACM-Reference-Format}
\bibliography{arxiv}
\end{CJK*}
\end{document}